\documentclass{article}
\usepackage{fancyvrb, xcolor}
\usepackage{spverbatim}
\usepackage{PRIMEarxiv}
\usepackage{float}
\usepackage[utf8]{inputenc} 
\usepackage[T1]{fontenc}    
\usepackage{hyperref}       
\usepackage{url}            
\usepackage{booktabs}       
\usepackage{amsfonts}       
\usepackage{nicefrac}       
\usepackage{microtype}      
\usepackage{lipsum}
\usepackage{fancyhdr}       
\usepackage{graphicx}       
\graphicspath{{media/}}     
\usepackage{amsmath}
\title{Flaws in the LLM Automation Narrative 
\thanks{\textit{\underline{Citation}}: 
\textbf{Perrett, G., Elliott, J., Hill, J., \& Scott, M. (2026). Flaws in the LLM Automation Narrative. }} 
}

\author{
  George Perrett\\
  New York University \\
  \texttt{gp77@nyu.edu} \\
   \And
  Javae Elliott\\
  New York University \\
  \AND
  Jennifer Hill \\
  New York University \\
  \texttt{jennifer.hill@nyu.edu} \\
  \And
  Marc Scott \\
  New York University \\
  \texttt{marc.scott@nyu.edu} \\
}

\begin{document}
\maketitle

\begin{abstract}
Large Language Models (LLMs) are increasingly described as performing at the level of human experts on knowledge economy tasks. These claims are primarily based on how LLMs perform on benchmarking tasks that measure average performance across standardized datasets. Primary limitations of many benchmarking tasks are that they often measure performance based on content directly included in LLM training data, and they frequently do not assess the reliability of LLM performance or the magnitude of LLM errors. However, in high stakes contexts, these qualities are critically important. Through a novel LLM benchmarking task that requires writing computer code to complete a data analysis task, we compare the performance of a frontier LLM against submissions from human experts and explicitly measure the variance of responses and the magnitude of errors. Our study reveals that the human experts perform better on average on a range of metrics and demonstrate less variability in performance. Our results provide evidence that LLMs do not consistently perform at the level of human experts and demonstrate the importance of measuring variance and assessing error magnitude in LLM benchmark evaluations. 

\end{abstract}


\section{Introduction}
The rapid rise of Large Language Models (LLMs), transformer-based machine learning models consisting of billions of trained parameters \cite{vaswani2017attention}, colloquially referred to as artificial intelligence (AI), has raised substantial economic, technical, and social questions \cite{hao2025empire, bender2021dangers}. Due to their size and power demands, the development and deployment of these models carry substantial economic costs. Current projections estimate that LLMs will require a cumulative investment of 7 trillion dollars by 2033 \cite{noffsinger2025cost}, and big tech companies have invested \$660 billion in the year of 2026 alone \cite{ft_capex}.

The developers of LLMs justify the substantial cost of developing and deploying them by promising that they will greatly augment or replace human expertise and can be reliably deployed across a variety of economically valuable tasks. Recent research from Anthropic, a prominent LLM developer, claims that currently available LLMs can \textit{theoretically} replace up to 94\% of computer and math work and around 90\% of work in finance, management, law, and administration \cite{massenkoff2026labor}. OpenAI, an LLM firm with a market valuation of \$840 billion \cite{ft_valuation}, has made similarly strong assertions, implying that ChatGPT (OpenAI's LLM) outperforms knowledge workers the majority of the time. As evidence, OpenAI cites ChatGPT's performance on their internally developed GDPVal benchmarking dataset \cite{patwardhan2025gdpval}. This benchmark compares work samples from knowledge economy workers with submissions from ChatGPT. Currently, ChatGPT 5.2 performs at or above the level of human workers 74.9\% of the time \emph{on GDPVal questions}. Describing ChatGPT as having the intelligence of "a legitimate PhD level expert in anything" \cite{gpt5, bbc}, OpenAI is not limiting these claims to their internal benchmarking tasks or even the set of all "ordinary" tasks but rather they are presenting LLMs as the equivalent of human experts. 

Benchmark datasets, such as GDPVal, are the most widely used approach for evaluating LLM capabilities in the academic and technical literature. Popular benchmarking tasks measure the performance of LLMs by assessing the proportion of questions or tasks an LLM can correctly complete \cite{zellers2019hellaswag, jimenez2024swe, wang2024mmlu, guha2023legalbench}. We argue that there are numerous limitations in the existing LLM benchmarking paradigm, and that these limitations have exaggerated the capabilities and reliability of LLMs.

\subsection{Limitations of Existing LLM Benchmark Evaluations}
There are several limitations of the existing LLM benchmark evaluation paradigm. First, and perhaps most fundamental, is the risk of benchmark contamination, where benchmark datasets, and the accompanying answers, are included in an LLM's training data \cite{white2024livebench, zhou2023don, balloccu2024leak}. As an analogy, a student's high score on a standardized test, such as the MCAT, GRE, or SAT, becomes considerably less impressive if we were told the student had a copy of the test's answer key while taking the exam. We make the same error when we confuse high performance on benchmark datasets with general intelligence or expertise. When measures are taken to ensure benchmark questions and answers are not included in the model's training data, performance rapidly deteriorates\cite{zhang2024careful}. A recent benchmark dataset, explicitly designed to ensure that questions and answers were not included in LLM training data, found that frontier LLMs could only answer 10\% of questions correctly \cite{center2026benchmark}. On the same benchmark, human experts answered 90\% of questions correctly. Evaluations of ChatGPT's performance on doctoral level biology exams reveal a similar pattern. While ChatGPT was able to answer some questions, its performance rapidly degraded on questions that are harder to represent in LLM training data \cite{kwong2026designing}. 

A second concern is the stochasticity of LLM responses, which is an inherent feature of how LLMs generate output \cite{banerjee2025llms}. Even when the temperature parameter, a setting that influences the stochasticity of responses, is set to low values, output from LLMs will still substantially vary across instantiations, and an LLM's ability to produce correct output in one instance does not ensure that future output will be correct \cite{vendrow2025large, barrie2024prompt}. However, the variation in LLM output is frequently overlooked when measuring the performance of LLMs. Current benchmarking approaches assess LLMs on many conceptually related but different questions, but often do not report the extent to which an LLM can consistently correctly answer \textit{the same} question or correctly repeat \textit{the same} task \cite{cao2025should}. 

An additional limitation of current LLM evaluations is the use of strictly binary performance criteria. Current benchmark approaches measure whether a model answered a question correctly or incorrectly, then compute the proportion of correct responses (accuracy), but ignore the magnitude of incorrect outputs. In actual deployment settings, the magnitude of errors is critically important. For example, in a complex computer coding task, deleting a production database, something that LLM coding agents have done \cite{mansoor2026claude}, is far more consequential than lesser errors like misspelling the name of a command. Understanding the frequency with which LLMs may or may not make catastrophic errors is an essential aspect of assessing their capacity, particularly as substitutes for experts. Because LLMs are unable to understand risk, cannot reason, and extrapolate poorly beyond the data the model was trained on \cite{bender2020climbing, shojaee2025illusion, barkan2025large}, they may exhibit a greater propensity for catastrophic errors, but this cannot be documented under current benchmarking procedures that do not measure the magnitude of LLM errors. 

Benchmark contamination, the failure to measure variance in response quality, and the failure to measure the propensity to catastrophic errors may help to explain the observed discrepancy between successful LLM benchmarks and unsuccessful LLM deployments in industry. While LLMs have performed astonishingly well on benchmark dataset within medical \cite{jin2021disease}, legal \cite{guha2023legalbench, katz2024gpt}, finance \cite{patwardhan2025gdpval, xie2024finben}, and computer programming \cite{jain2024livecodebench} contexts, they have performed poorly in actual externally valid medical \cite{bean2026reliability, wang2025scores, shool2025systematic, belisle2024we}, legal \cite{dahl2024large, magesh2025hallucination}, financial \cite{chen2025standard, bigeard2025finance}, and computer programming \cite{miserendino2025swe,noever2025can,becker2025measuring, jimenez2023swe} deployments. In fact, a study of pilot programs rolling out LLMs within industry found that, in 95\% of cases, the introduction of LLMs produced no return on investment \cite{mit}.


The present study aims to explain the discrepancy between the theoretical capabilities of LLMs documented in benchmarking evaluations and their actual performance in practice by evaluating LLM performance in a setting where we can directly measure the magnitude of errors made by an LLM, and we can assess the variation and reliability of LLM responses. As a secondary outcome, we are interested in directly testing the claim repeatedly made by prominent LLM developers that existing LLMs provide the equivalent of human expertise. In the following section, we provide a detailed explanation of our methodology and research design.

\section{Methods}

Our study asks how ChatGPT Codex 5.2 would have performed in the 2016 American Causal Inference Conference (ACIC) data competition. We leverage the datasets associated with the challenge to create a novel LLM benchmarking task that addresses the limitations of popular benchmarks. This setting allows us to measure the magnitude errors of tasks submitted by LLMs and is more robust against benchmark contamination. Moreover, we have designed our study to explicitly measure the variance of performance of an LLM on the same task. Another benefit of the 2016 American Causal Inference Data Challenge is that we have a natural comparison group of human PhD level experts, allowing us to explicitly test whether LLMs are, in fact, operating at the level of PhD experts. 

We have chosen to evaluate only the performance of the OpenAI ChatGPT Codex 5.2 LLM model because OpenAI is the only LLM developer to explicitly claim that its models have the intelligence of PhD level experts. We specifically selected the ChatGPT Codex 5.2 model because it was the most advanced model available at the time we performed our benchmark and is explicitly optimized for the programming and statistical tasks included in our benchmarking study\footnote{pilot testing confirmed that ChatGPT Codex 5.2 performed better than results from ChatGPT 5.2 Pro.}. In the following section, we present a detailed outline of our methodology.

\subsection{The 2016 ACIC Data Competition and Experimental Design}
The 2016 American Causal Inference Conference (ACIC) data competition (henceforth referred to as the competition) tasked teams of statisticians with writing a single executable script to analyze 7,700 unique datasets. Automated analysis with a single script was possible because, while each dataset was unique, they all shared a similar structure. Crucially, contestants lacked direct access to the datasets. Instead, they relied on a detailed set of instructions (a prompt) outlining the structure of the datasets, the analysis objectives, and the evaluation criteria. 
Moreover, the adversarial nature of the competition incentivized participants to keep their code and submissions private. Finally, unlike other popular data analysis competitions \cite{iglovikov2017satellite, bojer2021kaggle}, the participants' code was not publicly released. These aspects of the competition reduce the risk of benchmark contamination, making it a more reliable indicator of how models like ChatGPT Codex 5.2 would perform on tasks not directly included in the training data.

The original purpose of the competition was to compare methodological approaches for analyzing observational studies. A critical challenge in observational studies, a core subdiscipline within the field of causal inference, is that the treatment of interest has not been randomly assigned. To create a testing grounds for existing methods, the developers of the competition generated datasets from hypothetical observational studies with this feature. 
To help calibrate the analysis towards real-world challenges, each dataset consisted of an identical set of 58 predictor variables obtained from a real observational study. However, to ensure that critical assumptions were met and the true treatments effects were known, the outcome and binary treatment variables were synthetically generated (simulated) by the organizers of the competition and varied across datasets. Critically the treatment variable and the "potential outcomes" (outcomes that would manifest for each observation under each of the treatment regimes \cite{gelman2021regression})) were generated based only on the observed covariates. Therefore while the true treatment effects were known only to organizers, competition participants could plausibly obtain reasonable estimates of these effects from the observational data.\footnote{Refer to Dorie and colleagues \cite{dorie2019automated} for a more detailed description of the competition and the data generation processes used to create each dataset.} 
The primary advantage of synthetically generating (simulating) the treatment and outcome variables is that the competition's organizers had access to a `ground truth' of the true treatment effect
for each dataset, to objectively determine the best-performing submissions.


Competition participants submitted executable scripts that the organizers used to produce an estimate of the sample average treatment effect, $SATT$,  
and accompanying 95\% uncertainty intervals for each of the 7,700 datasets. Originally, there were 9 submissions from teams of PhD level statisticians (\emph{human experts (PhD)})\footnote{While we assume that entries were submitted because participants genuinely thought the method would do well, we can't rule out submissions made for the sake of exposing problems with a given method.}. An additional six submissions represented naive or older methods that serve as historical benchmarks of how well overly simplistic approaches would perform (\emph{historical strawmen}). 
Following the conclusion of the competition, the organizers created an additional nine submissions that combined features of the best performing methods (\emph{human experts (PhD) post-hoc}). 
In the ten years since the competition, aspects of the post-hoc methods, such as running algorithms with multiple chains, have become commonplace and recognized as best practices \cite{hill2023machine}.

To assess the performance of ChatGPT Codex 5.2, in the context of this competition we generated 20 independent executable scripts from ChatGPT Codex 5.2. Each of the 20 scripts was used to analyze the full set of 7,700 datasets. As a prompt, we provided the same detailed instructions that were given to the human participants in the competition. The full prompt is included in Appendix A. 



\subsection{Performance Measures}
Performance in the competition was assessed using four metrics: root mean squared error (RMSE), standardized bias, and 95\% interval coverage and length. 
Root mean squared error (RMSE), a common metric used in machine learning and statistical research, reflects the average magnitude of each submission's errors. Within the context of this data analysis task, RMSE is defined as 

$$\mathrm{RMSE} \equiv\sqrt{\frac{1}{7700}\sum_{i=1}^{7700} \bigg(\frac{\widehat{SATT_i} - SATT_i}{\sigma^y_{i}}\bigg)^2}$$

where $i$ is the $i$th dataset of the total 7,700 datasets,
$\widehat{SATT}$ is the estimated value of the true $SATT$ for each dataset. We divide by the standard deviation of the outcome variable $y$ in the $i$th dataset to account for differences in the outcome variable's scale across datasets. 

A related measure of distance between the estimates and their targets is bias. We measure bias across all datasets as

$$\text{standardized bias} \equiv \frac{1}{7700}\sum_{i=1}^{7700}\bigg(\frac{\widehat{SATT_i} - SATT_i}{\sigma^y_{i}}\bigg)$$

and again standardize by the standard deviation of the outcome. 
Unlike RMSE, standardized bias provides information on whether results are consistently over or underestimating the correct value. 

Interval coverage and length reflect the ability of a given approach to capture the uncertainty of the treatment effect estimate it yields. 
In the notation below, $\mathbb{I}$ is a binary indicator function\footnote{1 is true, 0 if false.}, and lci and uci are the upper and lower bounds of the 95\% uncertainty intervals respectively.  Interval coverage is measured as the proportion of 95\% uncertainty intervals that contain the corresponding true value of the $SATT$ and thus is defined as

$$\text{Coverage} = \frac{1}{7700}\sum_{i=1}^{7700}\mathbb{I} (\text{lci}_i \leq SATT_i \leq \text{uci}_i)$$

A proportion of .95 signals that the 95\% uncertainty intervals have been well calibrated. Lower proportions signal increasingly poor performance. Proportions greater than .95 are non-ideal but, in practice, are less concerning than coverage rates below .95. 

Because coverage only ensures that uncertainty intervals are wide enough to include the true value, it is important to consider interval length as well. In practice, we not only want intervals with coverage of .95 but also with small interval lengths. As with our other measures, we divide the interval length by the standard deviation of the outcome variable $y$ to control for differences in the scale of $ y$ across datasets. Formally, we define average interval length as: 

$$\text{interval length} = \frac{1}{7700}\sum_{i=1}^{7700}\frac{\text{uci}_i - \text{lci}_i}{\sigma_i^y}$$


Three of our four performance metrics (RMSE, standardized bias, and interval length) are continuous measures of performance. Unlike conventional benchmark studies that primarily rely on binary outcomes, our use of continuous measures allows us to assess the magnitude of ChatGPT's errors relative to human experts and explicitly measure the variance of output from ChatGPT Codex 5.2. In the following section, we present results from ChatGPT Codex 5.2 and compare its performance with that of human experts. 

\section{Results}

Of the 20 executable scripts generated by ChatGPT Codex 5.2, three (15\%) failed to run due to coding errors. Only the 17 executable scripts that can run are included in the subsequent results. All 20 scripts generated by ChatGPT Codex 5.2 are included in Appendix B. Generally, ChatGPT Codex 5.2 relied on two approaches to the problem. Eleven of the submissions from ChatGPT Codex 5.2 (55\%) attempted to implement different variations of Inverse Propensity Treatment Weighting (IPTW) or Augmented Inverse Propensity Treatment Weighting (AIPTW), and eight submissions from ChatGPT Codex 5.2 (40\%) implemented generalized causal forests \cite{athey2019estimating, wager2018estimation}. Notably, no submissions from ChatGPT Codex 5.2 attempted to implement Bayesian Additive Regression Trees (BART), a class of methods that has consistently been a top performer for the task described in the prompt \cite{hill2011bayesian, hahn2020bayesian, thal2023causal}.

Figure 1 compares the RMSE between submissions generated by human experts and ChatGPT Codex 5.2. The left panel of Figure 1 shows the RMSE for all 17 submissions from ChatGPT Codex 5.2 that ran, but our ability to visually compare performance is compromised by two catastrophically large RMSE values from the first and second scripts generated by ChatGPT Codex 5.2. The RMSE for both these scripts exceeds 100 billion standard deviations of the outcome variable. For readers less fluent in statistical terminology, .8 standard deviations of the outcome variable is considered a large effect size\cite{cohen2013applied}. In light of this reference point, an RMSE exceeding 100 billion is an error of an astounding magnitude. The eleventh, twelfth, fourteenth, and ninetieth scripts generated by ChatGPT Codex 5.2 also produced extreme RMSE values. The right panel of Figure 1 excludes these submissions to facilitate a clearer visual comparison of the remaining results. An evaluation of the code that produced extreme RMSE values revealed that these errors were due to subtle yet critical conceptual errors such as ignoring instructions about the datasets' underlying functional form, reweighting the sample incorrectly, or adding additional weight to observations with the most estimation error. Ultimately, these mistakes substantially amplifyied errors in the analysis.

\begin{figure}[h]
  \centering
  \includegraphics[width=5in]{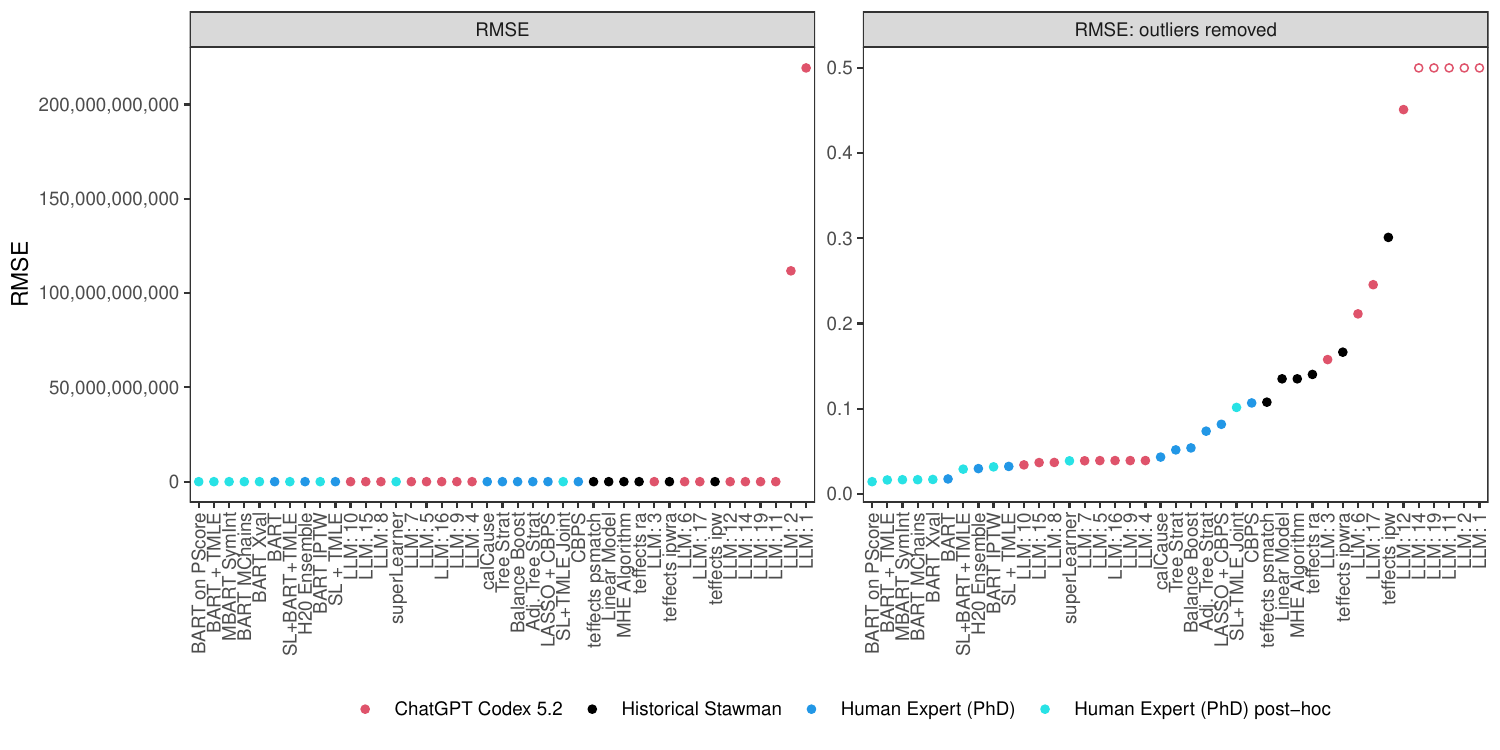}
  \caption{The RMSE among all submissions. The x-axis is ordered from smallest to largest RMSE value. Submissions from human experts are shown in blue, each submission from ChatGPT Codex 5.2 are shown in red, and historical strawman are shown in black. The right panel removes extreme values from 5 LLM submissions. The RMSE of ChatGPT Codex 5.2 submissions 14, 19, 11, 2, and 1 were 3.07,  1160.05, 2572.96, 111,765,692,519, and 219,316,810,584, respectively.}
  \label{fig:fig1}
\end{figure}

The majority of submissions from ChatGPT Codex 5.2 (red dots) performed notably worse than submissions from human experts (blue dots). A subset of eight submissions from ChatGPT Codex 5.2 performed reasonably well and produced RMSE values comparable to middle-tier performers in the original competitions. However, all scripts produced by ChatGPT Codex 5.2 fell short of the top performers in the competition, even though submissions were based on the state-of-the-art ten years ago. 

Figure 2 evaluates performance based on standardized bias. Again, results are obscured by exceedingly large values from poorly performing ChatGPT Codex 5.2 submissions. The five most extreme submissions from ChatGPT Codex 5.2 exhibit thousands of standard deviations of bias, with two submissions exceeding 1 billion standard deviations. The left panel shows all results, and the right panel shows results after removing the five most extreme submissions generated by ChatGPT Codex 5.2. 

\begin{figure}[h]
  \centering
  \includegraphics[width=5in]{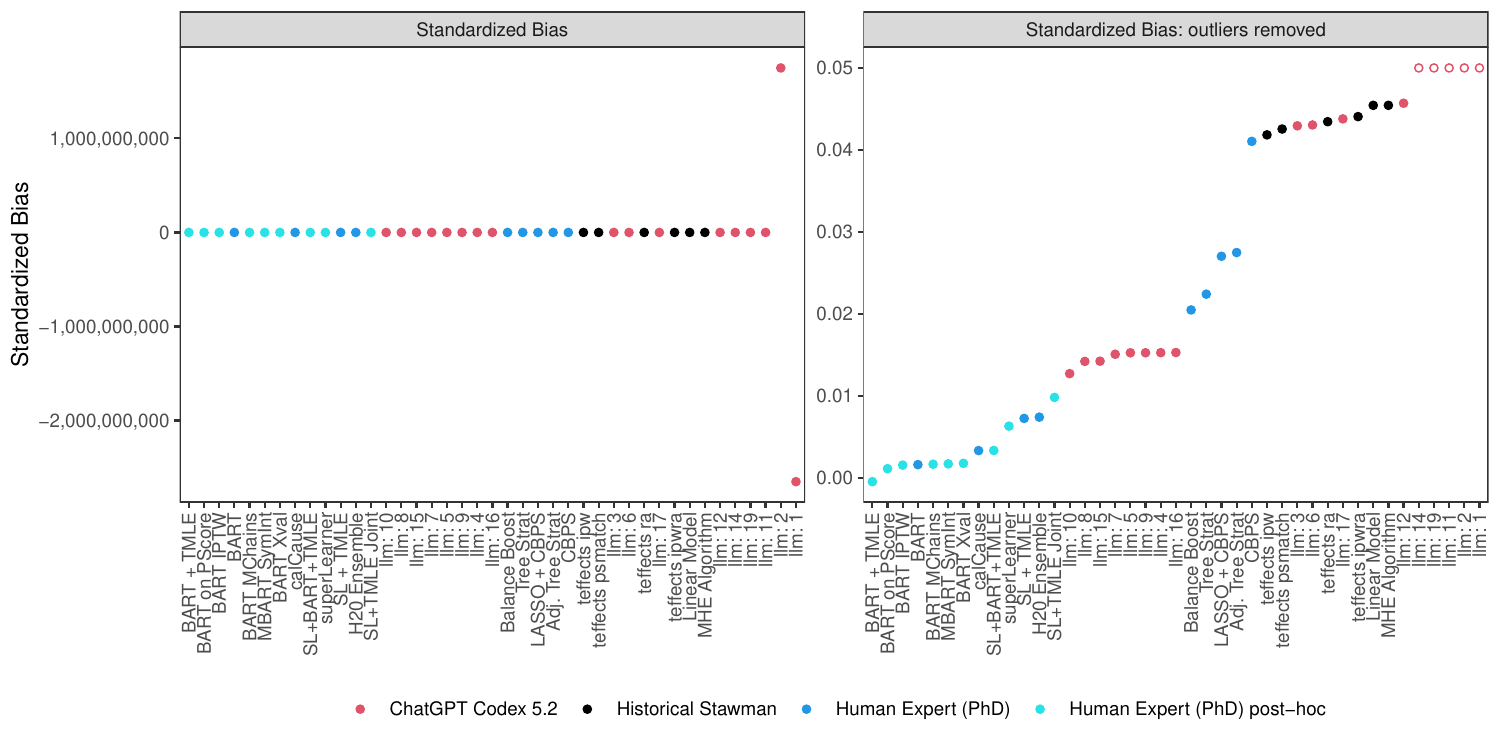}
  \caption{The standardized bias among all submissions. The x-axis is ordered from smallest to largest standardized bias value. Submissions from human experts are shown in blue, each submission from ChatGPT Codex 5.2 is shown in red, and historical strawman are shown in black. The right panel removes extreme values from 5 LLM submissions. The standardized bias of ChatGPT Codex 5.2 submissions 14, 19, 11, 2, and 1 were -2.8, 4.79, 30.83, 1,748,636,161, and -2,649,373,784.}
  \label{fig:fig2}
\end{figure}

Standardized bias tells a similar story to that of RMSE. The majority of submissions from ChatGPT Codex 5.2 are substantially worse than those submitted in the competition. The subset of submissions from ChatGPT Codex 5.2 that implemented causal forests outperformed the lower end human performers but still underperform relative to the majority of submissions in the original competition in terms of standardized bias. 

Both RMSE and standardized bias evaluate the quality of point estimates but provide minimal information about the quality of the provided 95\% uncertainty intervals. In the original competition, many submissions, even those with desirable point estimate properties, struggled with uncertainty quantification. Figure 3 presents the evaluation of coverage and interval length. To allow for better visual comparisons, methods with wider interval lengths were truncated to 0.2, even though they had much larger values. Recall that, ideally, a submission is close to .95 while also minimizing interval length. 

\begin{figure}[h]
  \centering
  \includegraphics[width=5in]{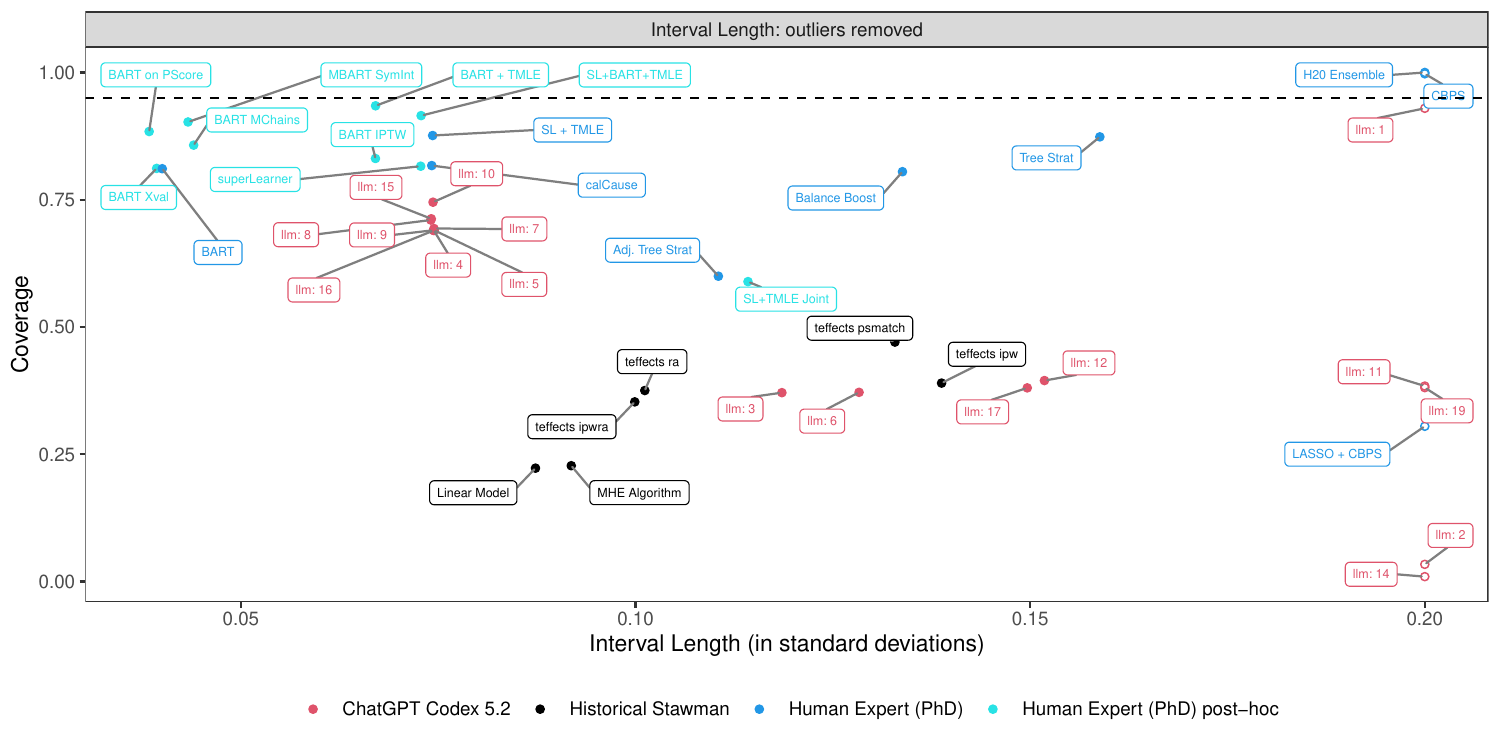}
  \caption{The coverage (y-axis) by interval length (x-axis) among all submissions. The x-axis is ordered from smallest to largest interval length. Coverage is averaged over all 7,700 datasets. Submissions from human experts are shown in blue, each submission from ChatGPT Codex 5.2 is shown in red, and historical strawman are shown in black. The interval lengths for 3 human and 5 ChatGPT Codex 5.2 submissions were removed. The average interval length for the human submission CBPS, LASSO+CBPS, and H20 Ensemble were .78, 4.97, and 6.08. The average interval length for ChatGPT Codex 5.2 submissions 14, 19, 11, 2, and 1 were 3.34, 294.83, 586.90, 234,300.74, 75,321,361,005.80.}
  \label{fig:fig3}
\end{figure}

Nearly all submissions from ChatGPT Codex 5.2 systematically undercover. While many of the submissions from human experts also undercover, generally they are closer to the ideal .95 and have noticeably shorter interval lengths. An important outlier is the first submission from ChatGPT Codex 5.2 (LLM: 1), which has a coverage rate close to .95. This becomes substantially less impressive and essentially disqualifies it as a successful contribution when we also consider that the interval length of this submission exceeded 65 billion standard deviations! 

While the eight ChatGPT Codex 5.2 submissions that implemented causal forests (LLM: 4,5,7,8,9,10,15,16) performed reasonably well on RMSE and standardized bias, they had noticeably lower coverage rates and simultaneously larger interval lengths than the majority of human expert submissions. 

Across all results, submissions from ChatGPT Codex 5.2 consistently performed worse than the majority of human submissions in the competition. Moreover, output from ChatGPT Codex 5.2 had alarmingly high variance. Among our continuous evolution criteria, RMSE, standardized bias, and interval length, submissions from human PhD level statisticians have a standard deviation of .029 for RMSE, a standard deviation of .012 for standardized bias, and a standard deviation of 1.76 for interval length. In sharp contrast, submissions from ChatGPT Codex 5.2 had standard deviations of 58171865734.93 for RMSE, 791722200.47 for standardized bias, and 18268109805.14 for interval length. The extreme variation in results from ChatGPT Codex 5.2 is not due to a single outlier but rather to five separate scripts that catastrophically fail to produce coherent results. Combining these five catastrophic scripts with the three non-executable scripts due to coding errors raises serious doubts about the reliability of ChatGPT Codex 5.2. 

We view the substantial variance in the performance of ChatGPT Codex 5.2 as a critical liability. Human experts participating in the competition submitted a rich variety of approaches to the problem, and yet, results from human participants are remarkably consistent. Even when considering the historical strawmen, methods that no causal inference statistician would seriously consider as a solution to the challenge, the variance in estimates from submissions by humans is small relative to the variance of those produced by Codex 5.2. 

Submissions from ChatGPT Codex 5.2 produced the opposite pattern. ChatGPT Codex 5.2 relied on two specific approaches, causal forests and IPTW/AIPTW, but results from these approaches were remarkably inconsistent. The pattern of our results clearly suggest that, on this task, asking ChatGPT Codex 5.2 to generate a strategy based on the prompt the humans received did not lead to results that were  consistently equivalent to human experts and exhibited substantial variation in quality.

\section{Discussion}

Our results may raise a series of immediate objections. 

A natural objection is the role of prompt engineering. It may be that with sufficient prompt engineering and scaffolding, it is possible to obtain better performance from ChatGPT Codex 5.2. However, the process of improving the prompt to produce better output would likely require substantial subject matter expertise in statistics and computer coding and would not reflect the capacity of the LLM but rather the expertise of the human writing the prompt. The fact that human submissions achieved relatively low variance and consistently produced quality work without additional prompt engineering but LLMs could not under the existing prompt is evidence that there remain real and distinct differences between the performance of human experts and LLMs.


Another objection is that we have presented results in an overly negative light and that in fact they confirm that ChatGPT Codex 5.2 has the \emph{capacity} to perform at the level of human experts. While a majority of submission from ChatGPT Codex 5.2 performed worse than human experts, including the numerous submission that failed to run, it is also true that eight of the twenty submissions (40\%) successfully used the \texttt{grf} R package to implement causal forests and produced results that were nearly as good as the best of the  human experts with respect to RMSE and bias (though more problematic with respect to uncertainty quantification). Critics could argue that we cannot disregard the notable subset of competitive submissions from ChatGPT Codex 5.2. However, even the best submissions from ChatGPT Codex 5.2 still perform worse than the best submissions from human experts. More importantly, we maintain that the variance of the results from ChatGPT Codex 5.2 is an extremely troubling finding. Consistency and reliability are necessary aspects of expertise and certainly pre-requisites for any automated workflows. 

In our results, ChatGPT Codex 5.2 submissions that implemented IPTW/AIPTW methods (LLM: 1, 2, 3, 6, 11, 12, 14, 17, and 19) performed particularly poorly. Across all implementations of IPTW/AIPTW ChatGPT Codex 5.2 attempted to implement the method from scratch without assistance from external (human written) R packages. Evaluation of the code produced by ChatGPT Codex 5.2 for these submission revealed consistent implementation errors of IPTW/AIPTW methods. Our results are not a criticism of IPTW/AIPTW on methodological grounds, but rather a criticism of ChatGPT Codex 5.2's ability to properly and consistently implement IPTW/AIPTW techniques. Annotations of these errors are included in Appendix B.

The discrepancy in performance between ChatGPT Codex 5.2 submissions that relied on the \texttt{grf} R package to implement causal forests and IPTW/AIPTW submissions, where ChatGPT Codex 5.2 attempted to implement the method directly, raises important questions. The \texttt{grf} R package is specifically designed to be easy to use and allow those not familiar with the technical details of causal forests to implement the underlying method. When the \texttt{grf} R package removed the difficult aspects of the task (implementing complex statistical methods), ChatGPT Codex 5.2 produced outputs comparable to some of the human submissions. In contrast, when ChatGPT attempted to complete the task without external assistance, the quality of submissions deteriorated. Implementing statistical methods is a difficult task that humans struggle with \cite{perrett2026scaffolding}. Our results suggest that delegating the task of implementing statistical methods to LLMs carries risks.



Our results show that while there is variation in submission quality among human experts, this variation is small. Recall that the standard deviation across human submissions was .029 for RMSE, .012 for standardized bias, and 1.76 for interval length. In contrast, the standard deviation for RMSE, standardized bias, and interval length from ChatGPT Codex 5.2 was in the billions. Even when the two largest outliers from ChatGTP Codex 5.2 are removed the standard deviation is 709 for RMSE, 8.67 for standardized bias, and 165 for interval length. The risk of LLMs is not that a model's output is never correct, but that \textit{a model's output is not consistently and reliably correct}. Our results show a pattern in which ChatGPT Codex 5.2 frequently makes errors of a catastrophic magnitude. It could be argued that, in practice, human workers would evaluate and check LLM produced code and these errors would never reach production. However, high profile failures, such as errors in LLM generated computer code that led to a 13 hour service disruption at AWS \cite{ft_aws}, a premier law firm submitting legal filings referencing hallucinated case law \cite{nyt-sullivan-cromwell-ai}, and recent researcher documenting that hallucinated citations are common across the published academic literature \cite{bauchner2026fabricated}, caution that this is likely not the case and align with our observed pattern of results. The extent to which humans can reliably identify LLM hallucinations remains an open question.

Substantively, our findings suggest that the performance of ChatGPT Codex 5.2 is not equivalent to that of our ensemble of PhD level experts. Moreover, the differences in performance are made all the more compelling by considering both the magnitude of errors and the variance in outputs within the same model, aspects a commonly overlooked in current LLM evaluations which typically only evaluate benchmark accuracy, the percentage of items an LLM can correctly answer. While benchmark accuracy is an important metric, ignoring the magnitude of errors occludes relevant information about how different LLMs differ from human experts. Likewise, this work implies that current benchmarking datasets fall short of capturing the variance of LLM output. Output from LLMs is inherently stochastic, and our results demontrate the degree to which results can vary from one run to the next. Understanding and documenting this variability is essential to fully comprehend the societal risks posed by the broader implementation of this technology.

\section*{Acknowledgments}
The authors gratefully acknowledge funding support from the Institute for Education Sciences grant number R305D240056.

\bibliographystyle{unsrt}  
\bibliography{references}  

\newpage
\section*{Appendix A: The prompt}

\begin{spverbatim}
Contest Motivation:

Causal inference researchers are constantly striving to create robust estimation procedures that will reliably estimate treatment effects across a wide variety of circumstances. This has led to a wide variety of methods that all purport to do be able to achieve this goal. However, typical papers in this field compare just two or three methods at a time. Moreover, these papers typically are written by researchers who, however well meaning, are interested in showcasing their own method. Thus it is unclear that such comparisons are entirely fair to the comparison methods considered. We would like to facilitate a broader comparison of methods in a setting in which the method considered is being implemented by someone who wants to show that method in its best light.

Therefore we are announcing a Causal Inference challenge, “Is your SATT where it’s at?” to better understand which approaches to causal inference perform well in particular observational study settings (described below) with point-in-time treatments. The goal is for individual researchers or research teams to obtain the best estimate of the treatment effect (specifically the effect of the treatment on the treated) for each dataset across a range of datasets.

In addition, there is increasing interest in developing methods for causal inference that are highly automated to decrease the burden on applied researchers, and yet produce accurate, precise, and reproducible estimates.

Structure of the Challenge:

77 datasets have been created by the organizers. The 58 available covariates in these datasets are drawn from a real study (to be revealed after the contest is over) and will be the same across the datasets. The binary treatment assignment and continuous outcome will be simulated for each dataset. The datasets will vary along the following dimensions:

- Level of nonlinearity (including discontinuities) of the assignment mechanism and response surface
- Level of treatment effect heterogeneity
- Ratio of treated to control observations
- Lack of overlap between treatment group and control group (there will be always be a reasonable amount of common support for the treatment group but there may exist controls in neighborhoods of the covariate space there no treated observations exist)
- Dimensionality of confounder space
- Magnitude of the treatment effect

All of the datasets will have the following features:

- Observations will be independent of each other and identically distributed (conditional on covariates)
- Ignorability (selection on observables, all confounders measured, no hidden bias…) will hold
- Not all covariates will be true confounders

All 77 simulation settings are used in this competition.

Create an executable R script.

Your executable R script should take two inputs, the name of data file and the name of an output file.

The data file will be in csv format and match the following specification:

- Column 1 is a binary treatment variable
- Column 2 is a continuous response variable
- Columns 3 and above are covariates; factors are coded with letters A/B/C/ binary variables are 0/1, and other columns are real numbers

Your output should consist of a csv file containing just the estimate of the treatment effect on the treated, a lower bound for a 95% confidence interval, and the corresponding upper bound.

(Optional) If you method is capable of providing individual effect estimates, those can be submitted as well. In this case, your executable should take three inputs: the input file, an output file for the estimated population effect, and an output file for the individual effects.

The format for the individual effects should also be a csv file with individual estimates, one per row in the same format as above.
\end{spverbatim}

\newpage
\section*{Appendix B: Annotated LLM code entries}

\subsection*{LLM: 1}

\begin{spverbatim}
args <- commandArgs(trailingOnly = TRUE)
infile <- args[1]
outfile <- args[2]

# Load data
dat <- read.csv(infile, stringsAsFactors = TRUE)

# Parse
T <- dat[[1]]
Y <- dat[[2]]
X <- dat[, -(1:2)]

# Convert factors to dummies
Xmat <- model.matrix(~ . -1, data = X)

\end{spverbatim}

\textcolor{red}{The instructions explicitly state that the outcome and treatment models are non-linear. Using a linear model for the propensity score will results in a biased estimate. The code then attempts to apply a double robust approach but fits another linear model as the outcome model. This is a critical problem because when both the treatment and outcome models are wrong a double robust solution will lead to larger errors. Given that the instructions explicitly say both the treatment and outcome models are non-linear this is a problem.}

\begin{spverbatim}
# Fit propensity model (logit)
ps_fit <- glm(T ~ Xmat, family = binomial())
ps <- predict(ps_fit, type = "response")

# Fit outcome regression model
y_fit <- glm(Y ~ T + Xmat)
mu1 <- predict(y_fit, newdata = data.frame(T = 1, Xmat))
mu0 <- predict(y_fit, newdata = data.frame(T = 0, Xmat))

\end{spverbatim}

\textcolor{red}{The most critical error is that the AIPTW technique is applied wrong 
correct code should be: \\
\texttt{treated <- which(T == 1) \\
 att <- mean(\\
 y[treated] - ((y[treated]*ps[treated] + mu0*(T - ps))/(1 - ps))\\
 )
 }
 }

\begin{spverbatim}
# AIPW estimator of SATT
treated <- which(T == 1)
att <- mean(
  (Y[treated] - mu0[treated]) -
    (ps[treated]/(1-ps[treated])) * (Y[treated] - mu1[treated])
)

\end{spverbatim}
\textcolor{red}{
The code below attempts to bootstrap standard errors but relying on 200 bootstrap samples is to small and will lead to inaccurate standard errors this should be set to at least 1000.
}
\begin{spverbatim}
# Bootstrap CI
B <- 200 
boot_vals <- numeric(B)
set.seed(123)
for (b in 1:B) {
  idx <- sample(1:nrow(dat), replace = TRUE)
  db <- dat[idx, ]
  T_b <- db[[1]]
  Y_b <- db[[2]]
  X_b <- model.matrix(~ . -1, data = db[, -(1:2)])
  
  ps_b <- glm(T_b ~ X_b, family = binomial()) |> 
    predict(type="response")
  y_b <- glm(Y_b ~ T_b + X_b)
  
  mu1_b <- predict(y_b, newdata = data.frame(T_b=1, X_b))
  mu0_b <- predict(y_b, newdata = data.frame(T_b=0, X_b))
  
  treated_b <- which(T_b==1)
  boot_vals[b] <- mean(
    (Y_b[treated_b] - mu0_b[treated_b]) -
      (ps_b[treated_b]/(1-ps_b[treated_b])) * (Y_b[treated_b] - 
                                                 mu1_b[treated_b])
  )
}

ci <- quantile(boot_vals, c(0.025, 0.975))

# Write output
out <- data.frame(estimate = att, CI_lower = ci[1], CI_upper = ci[2])
write.csv(out, outfile, row.names = FALSE)

\end{spverbatim}

\subsection*{LLM: 2}

\begin{spverbatim}

suppressMessages({
  library(stats)
})

# ----- Read command line arguments -----
args <- commandArgs(trailingOnly = TRUE)
if (length(args) < 2) {
  stop("Usage: Rscript estimate_satt.R input.csv output.csv")
}

infile  <- args[1]
outfile <- args[2]

# ----- Load data -----
dat <- read.csv(infile, stringsAsFactors = TRUE)
Tt <- dat[[1]]          # treatment
Y  <- dat[[2]]          # outcome
X  <- dat[, -c(1,2)]

# Convert factors to model matrix
Xmat <- model.matrix(~ . -1, data = X)
\end{spverbatim}
\textcolor{red}{The instructions explicitly state that the outcome and treatment models are non-linear. Using a linear model for the propensity score will results in a biased estimate. The code then attempts to apply a double robust approach but fits another linear model as the outcome model. This is a critical problem because when both the treatment and outcome models are wrong a double robust solution will lead to larger errors. Given that the instructions explicitly say both the treatment and outcome models are non-linear this is a problem.}
\begin{spverbatim}
# ----- Propensity score model -----
prop.mod <- glm(Tt ~ Xmat, family = binomial())
ps <- predict(prop.mod, type = "response")

# ----- Outcome model -----
out.mod <- lm(Y ~ Tt + Xmat)
mu1 <- predict(out.mod, newdata = data.frame(Tt = 1, Xmat))
mu0 <- predict(out.mod, newdata = data.frame(Tt = 0, Xmat))
\end{spverbatim}
\textcolor{red}{The most critical error is that the AIPTW technique is applied wrong 
correct code should be: \\
\texttt{treated <- which(T == 1) \\
 att <- mean(\\
 y[treated] - ((y[treated]*ps[treated] + mu0*(T - ps))/(1 - ps))\\
 )
 }
 }
\begin{spverbatim}
# ----- Doubly robust ATT estimator -----
# AIPW estimator for ATT
w <- Tt + (1 - Tt) * ps/(1-ps)
att_est <- mean( (Tt*(Y - mu0)) - ((1-Tt)*ps/(1-ps))*(Y - mu0) )
\end{spverbatim}
\textcolor{red}{
The code below attempts to bootstrap standard errors but relying on 200 bootstrap samples is to small and will lead to inaccurate standard errors this should be set to at least 1000.
}
\begin{spverbatim}
# ----- Bootstrap for CI -----
set.seed(1)
B <- 200
att_boot <- numeric(B)
n <- nrow(dat)

for (b in 1:B) {
  idx <- sample(1:n, replace = TRUE)
  Tt_b <- Tt[idx]
  Y_b  <- Y[idx]
  X_b  <- Xmat[idx, , drop=FALSE]
  
  prop.mod_b <- glm(Tt_b ~ X_b, family = binomial())
  ps_b <- predict(prop.mod_b, type="response")
  
  out.mod_b <- lm(Y_b ~ Tt_b + X_b)
  mu0_b <- predict(out.mod_b, newdata = data.frame(Tt_b=0, X_b))
  
  att_boot[b] <- mean( (Tt_b*(Y_b - mu0_b)) -
                         ((1-Tt_b)*ps_b/(1-ps_b))*(Y_b - mu0_b) )
}

ci <- quantile(att_boot, probs = c(0.025, 0.975))

# ----- Output -----
out <- data.frame(ATT = att_est, CI_L = ci[1], CI_U = ci[2])
write.csv(out, outfile, row.names = FALSE)
\end{spverbatim}

\subsection*{LLM: 3}
\begin{spverbatim}
args <- commandArgs(trailingOnly = TRUE)
if (length(args) < 2) {
  stop("Usage: Rscript estimate_satt.R input.csv output.csv")
}

infile  <- args[1]
outfile <- args[2]

dat <- read.csv(infile, stringsAsFactors = TRUE)

treat <- dat[,1]
y     <- dat[,2]
X     <- dat[,-c(1,2), drop=FALSE]

# Build model matrix (handles factors automatically)
Xmat <- model.matrix(~ ., data=X)
Xmat <- Xmat[,-1, drop=FALSE]  # remove intercept
\end{spverbatim}
\textcolor{red}{This submission correctly implements the IPTW weights but ignores that the instructions clearly stating that the assignment mechanism is non-linear. Using a linear model to estimate the propensity score will create problems}
\begin{spverbatim}
# Propensity score model
ps_mod <- glm(treat ~ Xmat, family=binomial)
ps <- predict(ps_mod, type="response")
ps <- pmin(pmax(ps, 0.01), 0.99)  # trim for stability

# Outcome model (controls only)
dat_ctrl <- data.frame(y=y[treat==0], Xmat[treat==0, , drop=FALSE])
out_mod <- lm(y ~ ., data=dat_ctrl)

# Predict m0(x)
m0 <- predict(out_mod, newdata=data.frame(Xmat))

# DR ATT estimator
w <- ps/(1-ps)
att <- mean(y[treat==1] - m0[treat==1]) -
  sum(w[treat==0] * (y[treat==0] - m0[treat==0])) / 
  sum(w[treat==0])
\end{spverbatim}
\textcolor{red}{
The code below attempts to bootstrap standard errors but relying on 200 bootstrap samples is to small and will lead to inaccurate standard errors this should be set to at least 1000.
}
\begin{spverbatim}
# Bootstrap CI
B <- 200
boot_est <- numeric(B)
set.seed(123)

for (b in 1:B) {
  idx <- sample(seq_len(nrow(dat)), replace=TRUE)
  tb <- treat[idx]
  yb <- y[idx]
  Xb <- X[idx, , drop=FALSE]
  Xbmat <- model.matrix(~ ., data=Xb)
  Xbmat <- Xbmat[,-1, drop=FALSE]
  
  ps_mod_b <- glm(tb ~ Xbmat, family=binomial)
  psb <- predict(ps_mod_b, type="response")
  psb <- pmin(pmax(psb, 0.01), 0.99)
  
  dat_ctrl_b <- data.frame(y=yb[tb==0], Xbmat[tb==0,,drop=FALSE])
  out_mod_b <- lm(y ~ ., data=dat_ctrl_b)
  m0b <- predict(out_mod_b, newdata=data.frame(Xbmat))

  wb <- psb/(1-psb)
  boot_est[b] <- mean(yb[tb==1] - m0b[tb==1]) -
    sum(wb[tb==0]*(yb[tb==0]-m0b[tb==0])) / 
    sum(wb[tb==0])
}

ci <- quantile(boot_est, probs=c(0.025,0.975), na.rm=TRUE)

# Write output
out <- data.frame(estimate=att, lower=ci[1], upper=ci[2])
write.csv(out, outfile, row.names=FALSE)
\end{spverbatim}

\subsection*{LLM: 4}
\begin{spverbatim}
args <- commandArgs(trailingOnly = TRUE)
input_file  <- args[1]
output_file <- args[2]

# load packages
suppressMessages(library(grf))

# read data
dat <- read.csv(input_file)

# extract treatment, outcome, covariates
W <- dat[[1]]
Y <- dat[[2]]
X <- dat[ , -c(1,2)]

# convert character columns to factors
X <- data.frame(lapply(X, function(x){
  if (is.character(x)) factor(x) else x
}))

# create model matrix for factors
X_mat <- model.matrix(~ . - 1, data = X)

# causal forest
cf <- causal_forest(X_mat, Y, W)

# estimate ATT / SATT
att <- average_treatment_effect(cf, target.sample = "treated")
est <- att["estimate"]
se  <- att["std.err"]

# 95% CI
lwr <- est - 1.96*se
upr <- est + 1.96*se

# output
out <- data.frame(estimate = est, lwr = lwr, upr = upr)
write.csv(out, output_file, row.names = FALSE)
\end{spverbatim}

\subsection*{LLM: 5}
\begin{spverbatim}
suppressPackageStartupMessages({
  if (!require("grf")) {
    stop("Package 'grf' is required. Install with 
install.packages('grf').")
  }
})

args <- commandArgs(trailingOnly = TRUE)
if (length(args) < 2) {
  stop("Usage: Rscript estimate_att.R input.csv output.csv 
[individual_effects.csv]")
}

input_file  <- args[1]
output_file <- args[2]
indiv_file  <- if (length(args) >= 3) args[3] else NULL

# ---- Read data ----
dat <- read.csv(input_file, stringsAsFactors = TRUE)

# Columns: 1 = treatment, 2 = outcome, 3+ covariates
W <- dat[[1]]
Y <- dat[[2]]
Xraw <- dat[, -c(1,2), drop=FALSE]

# Convert factors to numeric dummy matrix
X <- model.matrix(~ . - 1, data = Xraw)

# ---- Fit causal forest ----
set.seed(123)
cf <- causal_forest(X, Y, W)

# ---- Estimate ATT (SATT) ----
att <- average_treatment_effect(cf, target.sample = "treated")
est <- att["estimate"]
se  <- att["std.err"]

lower <- est - 1.96 * se
upper <- est + 1.96 * se

# ---- Write output ----
out <- data.frame(estimate = est, lower = lower, upper = upper)
write.csv(out, output_file, row.names = FALSE)

# ---- Optional individual effects ----
if (!is.null(indiv_file)) {
  cates <- predict(cf)\$predictions
  write.csv(data.frame(cate = cates), indiv_file, row.names = FALSE)
}
\end{spverbatim}

\subsection*{LLM: 6}
\begin{spverbatim}
args <- commandArgs(trailingOnly = TRUE)
if (length(args) < 2) {
  stop("Usage: Rscript estimate_satt.R input.csv output.csv")
}
infile <- args[1]
outfile <- args[2]

set.seed(123)

# ---- Read data ----
dat <- read.csv(infile, stringsAsFactors = TRUE)

# Expected columns:
# 1 = treatment (binary)
# 2 = outcome (continuous)
# 3+ = covariates

T <- dat[[1]]
Y <- dat[[2]]
X <- dat[, -(1:2), drop = FALSE]

# create model matrix for covariates (handle factors)
Xmat <- model.matrix(~ . , data = X)[, -1, drop = FALSE]

# ---- AIPW ATT estimator ----
att_aipw <- function(Y, T, Xmat) {
  # Propensity score
  ps_mod <- glm(T ~ Xmat, family = binomial())
  ehat <- pmin(pmax(predict(ps_mod, type = "response"), 0.01), 0.99)
  
  # Outcome model on controls only
  ym_mod <- lm(Y ~ Xmat, subset = (T == 0))
  mu0_hat <- predict(ym_mod, newdata = data.frame(Xmat))
  
  # AIPW ATT
  w <- T - (1 - T) * ehat / (1 - ehat)
  att <- mean((Y - mu0_hat) * w) / mean(T)
  return(att)
}

# point estimate
att_hat <- att_aipw(Y, T, Xmat)

# ---- Bootstrap CI ----
B <- 200
n <- nrow(dat)
boot <- numeric(B)

for (b in 1:B) {
  idx <- sample(1:n, n, replace = TRUE)
  boot[b] <- att_aipw(Y[idx], T[idx], Xmat[idx, , drop = FALSE])
}

ci <- quantile(boot, c(0.025, 0.975))

# ---- Write output ----
out <- data.frame(estimate = att_hat,
                  ci_lower = ci[1],
                  ci_upper = ci[2])

write.csv(out, file = outfile, row.names = FALSE)
\end{spverbatim}

\subsection*{LLM: 7}
\begin{spverbatim}
suppressMessages({
  library(grf)
})

args <- commandArgs(trailingOnly = TRUE)
if(length(args) < 2) {
  stop("Usage: Rscript my_method.R input.csv output.csv 
[individual_effects.csv]")
}

infile  <- args[1]
outfile <- args[2]
indfile <- ifelse(length(args) >= 3, args[3], NA)

# Read data
dat <- read.csv(infile, stringsAsFactors = TRUE)

W <- dat[[1]]   # treatment
Y <- dat[[2]]   # outcome
X <- dat[, -c(1,2)]

# Convert factors into numeric model matrix
Xmat <- model.matrix(~ . , data = X)[, -1]

# Fit causal forest
set.seed(123)
cf <- causal_forest(Xmat, Y, W)

# Estimate ATT (SATT)
att <- average_treatment_effect(cf, target.sample = "treated")
est <- att[["estimate"]]
se  <- att[["std.err"]]

lower <- est - 1.96 * se
upper <- est + 1.96 * se

# Write estimate output
out <- data.frame(estimate = est, lower = lower, upper = upper)
write.csv(out, outfile, row.names = FALSE)

# Optional: individual treatment effects
if(!is.na(indfile)) {
  tauhat <- predict(cf)\$predictions
  write.csv(data.frame(tau = tauhat), indfile, row.names = FALSE)
}
\end{spverbatim}

\subsection*{LLM: 8}
\begin{spverbatim}
args <- commandArgs(trailingOnly = TRUE)
if (length(args) < 2) {
  stop("Usage: Rscript run_att.R input.csv output.csv")
}
infile  <- args[1]
outfile <- args[2]

# ---- Load data ----
dat <- read.csv(infile, stringsAsFactors = TRUE)

W <- dat[,1]              # treatment (binary)
Y <- dat[,2]              # outcome (continuous)
X <- dat[,-c(1,2)]         # covariates

# Convert factors to dummy variables
X <- model.matrix(~ . -1, data = X)

# ---- Fit causal forest ----
suppressMessages(library(grf))

cf <- causal_forest(X, Y, W, tune.parameters = "all")

# ---- Estimate ATT (SATT) ----
att <- average_treatment_effect(cf, target.sample = "treated")
est <- att["estimate"]
se  <- att["std.err"]

ci.lower <- est - 1.96 * se
ci.upper <- est + 1.96 * se

# ---- Output results ----
out <- data.frame(estimate = est,
                  ci.lower = ci.lower,
                  ci.upper = ci.upper)

write.csv(out, outfile, row.names = FALSE)
\end{spverbatim}
\subsection*{LLM:9}
\begin{spverbatim}
#!/usr/bin/env Rscript

suppressPackageStartupMessages({
  library(grf)
})

args <- commandArgs(trailingOnly = TRUE)

if(length(args) < 2){
  stop("Usage: Rscript run_method.R input.csv output_effect.csv 
[output_indiv.csv]")
}

input_file  <- args[1]
output_file <- args[2]
indiv_file  <- ifelse(length(args) >= 3, args[3], NA)

# Read data
dat <- read.csv(input_file, stringsAsFactors = TRUE)

# Treatment & outcome
W <- dat[[1]]
Y <- dat[[2]]

# Covariates
X <- dat[, -c(1,2)]

# Convert factors to dummies
X <- model.matrix(~ . - 1, data = X)

# Fit causal forest
cf <- causal_forest(X, Y, W)

# Estimate ATT (SATT)
att <- average_treatment_effect(cf, target.sample = "treated")

# Output: estimate, lower 95% CI, upper 95% CI
out <- data.frame(
  estimate = att[["estimate"]],
  ci_lower = att[["estimate"]] - 1.96 * att[["std.err"]],
  ci_upper = att[["estimate"]] + 1.96 * att[["std.err"]]
)

write.csv(out, output_file, row.names = FALSE)

# Optional: individual treatment effects
if(!is.na(indiv_file)){
  tau_hat <- predict(cf)\$predictions
  write.csv(data.frame(ite = tau_hat), indiv_file, row.names = FALSE)
}
\end{spverbatim}
\subsection*{LLM: 10}
\begin{spverbatim}
#!/usr/bin/env Rscript

# Usage:
# Rscript satt_estimator.R input.csv output.csv [individual_output.csv]

args <- commandArgs(trailingOnly = TRUE)
if (length(args) < 2) stop("Need input file and output file")

infile  <- args[1]
outfile <- args[2]
indfile <- if (length(args) >= 3) args[3] else NA

# Load libraries
suppressMessages({
  library(grf)
})

# Read data
dat <- read.csv(infile)

# Treatment and outcome
W <- dat[[1]]
Y <- dat[[2]]

# Covariates
X <- dat[, -c(1,2), drop = FALSE]

# Convert character variables to factors
X <- data.frame(lapply(X, function(col) {
  if (is.character(col)) as.factor(col) else col
}))

# One‑hot encode factors
X_mat <- model.matrix(~ . - 1, data = X)

# Fit causal forest
set.seed(1)
cf <- causal_forest(X_mat, Y, W, num.trees = 2000,
                    tune.parameters = "all")

# Average treatment effect on treated (SATT)
ate <- average_treatment_effect(cf, target.sample = "treated", method
                                = "AIPW")

est <- ate["estimate"]
se  <- ate["std.err"]
ci  <- c(est - 1.96 * se, est + 1.96 * se)

# Output result
out <- data.frame(estimate = est, ci.lower = ci[1], ci.upper = ci[2])
write.csv(out, outfile, row.names = FALSE)

# Optional: individual treatment effects
if (!is.na(indfile)) {
  tauhat <- predict(cf)\$predictions
  write.csv(data.frame(tau = tauhat), indfile, row.names = FALSE)
}
\end{spverbatim}
\subsection*{LLM: 11}
\begin{spverbatim}
#!/usr/bin/env Rscript

args <- commandArgs(trailingOnly=TRUE)
if (length(args) < 2) {
  stop("Usage: Rscript submit.R input.csv output.csv")
}

infile  <- args[1]
outfile <- args[2]

dat <- read.csv(infile)

# Extract variables
treat <- dat[,1]
y     <- dat[,2]
x     <- dat[,-c(1,2)]

# Convert categorical columns to factors
x[] <- lapply(x, function(col){
  if (is.character(col)) as.factor(col) else col
})
\end{spverbatim}
\textcolor{red}{The instructions explicitly state that the outcome and treatment models are non-linear. Using a linear model for the propensity score will results in a biased estimate. The code then attempts to apply a double robust approach but fits another linear model as the outcome model. This is a critical problem because when both the treatment and outcome models are wrong a double robust solution will lead to larger errors. Given that the instructions explicitly say both the treatment and outcome models are non-linear this is a problem.}
\begin{spverbatim}
# ---------------------------
# 1. Propensity score model
# ---------------------------
ps_mod <- glm(treat ~ ., data=cbind(treat, x), family=binomial)
ps_hat <- predict(ps_mod, type="response")

# ---------------------------
# 2. Outcome regression model
# ---------------------------
y_mod <- lm(y ~ treat + ., data=cbind(y, treat, x))
y1_hat <- predict(y_mod, newdata=cbind(treat=1, x))
y0_hat <- predict(y_mod, newdata=cbind(treat=0, x))
\end{spverbatim}
\textcolor{red}{The most critical problem is the AIPTW weighting technique is applied incorrectly. Dividing by proportion of treated units is incorrect. This code is not an AIPTW estimator and will produce biased estimates even when the outcome model is correctly specified.}
\begin{spverbatim}
# ---------------------------
# 3. Doubly Robust ATT
# ---------------------------
w <- treat + (1 - treat) * ps_hat/(1-ps_hat)
att_hat <- mean((treat * (y - y0_hat)) / mean(treat)) -
  mean(((1 - treat) * ps_hat/(1-ps_hat) * (y - y0_hat)) / 
         mean(treat))
\end{spverbatim}
\textcolor{red}{
The code below attempts to bootstrap standard errors but relying on 200 bootstrap samples is to small and will lead to inaccurate standard errors this should be set to at least 1000.
}
\begin{spverbatim}
# ---------------------------
# 4. Bootstrap CI
# ---------------------------
B <- 200
boot_att <- numeric(B)

set.seed(123)
n <- nrow(dat)
for (b in 1:B){
  idx <- sample(1:n, replace=TRUE)
  dat_b <- dat[idx,]
  treat_b <- dat_b[,1]
  y_b <- dat_b[,2]
  x_b <- dat_b[,-c(1,2)]
  x_b[] <- lapply(x_b, function(col){
    if (is.character(col)) as.factor(col) else col
  })
  
  ps_mod_b <- glm(treat_b ~ ., data=cbind(treat_b, x_b), 
                  family=binomial)
  ps_hat_b <- predict(ps_mod_b, type="response")
  
  y_mod_b <- lm(y_b ~ treat_b + ., data=cbind(y_b, treat_b, x_b))
  y0_hat_b <- predict(y_mod_b, newdata=cbind(treat_b=0, x_b))
  
  boot_att[b] <- mean((treat_b * (y_b - y0_hat_b)) / mean(treat_b)) -
    mean(((1 - treat_b) * ps_hat_b/(1-ps_hat_b) * (y_b -
                                                     y0_hat_b)) / mean(treat_b))
}

ci <- quantile(boot_att, probs=c(0.025, 0.975))

# ---------------------------
# 5. Output
# ---------------------------
out <- data.frame(att=att_hat, lower=ci[1], upper=ci[2])
write.csv(out, outfile, row.names=FALSE)
\end{spverbatim}

\subsection*{LLM: 12}
\begin{spverbatim}
#!/usr/bin/env Rscript

args <- commandArgs(trailingOnly = TRUE)
if (length(args) < 2) {
  stop("Usage: script.R input.csv output.csv")
}

input_file  <- args[1]
output_file <- args[2]

dat <- read.csv(input_file, stringsAsFactors = TRUE)

# Required columns: 1 = treatment, 2 = outcome, 3+ = covariates
T <- dat[[1]]
Y <- dat[[2]]
X <- dat[, 3:ncol(dat), drop = FALSE]

# Build model matrix (handles factors)
X_mat <- model.matrix(~ . , data = X)[, -1, drop=FALSE]
\end{spverbatim}
\textcolor{red}{This submission properly implements IPTW, however, it ignored that the instructions explicitly stated that the assignment mechanism is non-linear. Fitting a linear model to estimate the propensity score is incorrect.}
\begin{spverbatim}
# Propensity score model
ps_model <- glm(T ~ X_mat, family = binomial)
ps <- predict(ps_model, type = "response")

# ATT weights: treated weight=1, control weight=ps/(1-ps)
w <- ifelse(T == 1, 1, ps / (1 - ps))

# ATT estimate
att_hat <- weighted.mean(Y[T==1], w[T==1]) - weighted.mean(Y[T==0], w[T==0])
\end{spverbatim}
\textcolor{red}{
The code below attempts to bootstrap standard errors but relying on 200 bootstrap samples is to small and will lead to inaccurate standard errors this should be set to at least 1000.
}
\begin{spverbatim}
# Bootstrap CI
B <- 200
boot_att <- numeric(B)
set.seed(123)
n <- length(T)
for (b in 1:B) {
  idx <- sample(1:n, n, replace = TRUE)
  Tb <- T[idx]; Yb <- Y[idx]; Xb <- X[idx, , drop=FALSE]
  Xb_mat <- model.matrix(~ . , data=Xb)[, -1, drop=FALSE]
  ps_b <- predict(glm(Tb ~ Xb_mat, family=binomial), type="response")
  w_b <- ifelse(Tb==1,1, ps_b/(1-ps_b))
  boot_att[b] <- weighted.mean(Yb[Tb==1], w_b[Tb==1]) -
    weighted.mean(Yb[Tb==0], w_b[Tb==0])
}
ci <- quantile(boot_att, probs=c(0.025,0.975))

# Output
out <- data.frame(estimate = att_hat,
                  ci_lower = ci[1],
                  ci_upper = ci[2])

write.csv(out, output_file, row.names = FALSE)
\end{spverbatim}
\subsection*{LLM: 13}
\begin{spverbatim}
#!/usr/bin/env Rscript

args <- commandArgs(trailingOnly=TRUE)

if (length(args) < 2) {
  stop("Usage: Rscript estimate_att.R input.csv output.csv")
}

input_file <- args[1]
output_file <- args[2]

# Load data
dat <- read.csv(input_file, stringsAsFactors = TRUE)

# Treatment, outcome, covariates
W <- dat[,1]
Y <- dat[,2]
X <- dat[,-c(1,2)]

# Propensity score model (logistic regression)
ps_model <- glm(W ~ ., data = data.frame(W=W, X), family = binomial)
ps <- predict(ps_model, type="response")

# Outcome model (separate models for treated/control)
dat_treated <- data.frame(Y=Y[W==1], X=X[W==1,])
dat_control <- data.frame(Y=Y[W==0], X=X[W==0,])
\end{spverbatim}
\textcolor{red}{
There is an error in the code below and the script will not run. The code is attempting to apply a categorical variables not in the training data. This will produce an error. Moreover, the error comes from tying to fit a model that is not necessary for the AIPTW method that Codex is attempting to implement. The object m1 is what is creating the issue, but this object is not necessary to estimate the ATT... Notice that mu1 is never actually used. This is a hallucination. 
}
\begin{spverbatim}
outcome_treated <- lm(Y ~ ., data=dat_treated)
outcome_control <- lm(Y ~ ., data=dat_control)

mu1 <- predict(outcome_treated, newdata=data.frame(X=X))
mu0 <- predict(outcome_control, newdata=data.frame(X=X))

# Doubly robust ATT estimator (AIPW)
att_scores <- (W*(Y - mu0)) - ((1-W)*ps/(1-ps))*(Y - mu0)
att <- mean(att_scores[W==1])

# Bootstrap CI
set.seed(123)
B <- 200
boot_att <- numeric(B)
n <- nrow(dat)

for (b in 1:B) {
  idx <- sample(1:n, replace=TRUE)
  Wb <- W[idx]
  Yb <- Y[idx]
  Xb <- X[idx,]
  
  psb <- glm(Wb ~ ., data=data.frame(Wb=Wb, Xb), family=binomial)
  ps_hat <- predict(psb, type="response")
  
  out1 <- lm(Yb[Wb==1] ~ ., data=data.frame(Yb=Yb[Wb==1], 
                                            Xb=Xb[Wb==1,]))
  out0 <- lm(Yb[Wb==0] ~ ., data=data.frame(Yb=Yb[Wb==0], 
                                            Xb=Xb[Wb==0,]))
  
  mu0b <- predict(out0, newdata=data.frame(Xb=Xb))
  
  scoreb <- (Wb*(Yb - mu0b)) - ((1-Wb)*ps_hat/(1-ps_hat))*(Yb - mu0b)
  boot_att[b] <- mean(scoreb[Wb==1])
}

ci <- quantile(boot_att, probs=c(0.025,0.975), na.rm=TRUE)

# Write output
out <- data.frame(estimate=att, ci_lower=ci[1],ci_upper=ci[2])
write.csv(out, output_file, row.names=FALSE)
\end{spverbatim}
\subsection*{LLM: 14}
\begin{spverbatim}
#!/usr/bin/env Rscript

# ---- Packages ----
pkgs <- c("ranger", "data.table")
for (p in pkgs) {
  if (!requireNamespace(p, quietly = TRUE)) {
    install.packages(p, repos = "https://cloud.r-project.org")
  }
}
library(ranger)
library(data.table)

# ---- Read arguments ----
args <- commandArgs(trailingOnly = TRUE)
if (length(args) < 2) {
  stop("Usage: Rscript satt_estimator.R input.csv output.csv")
}
infile <- args[1]
outfile <- args[2]

# ---- Load data ----
dat <- fread(infile, data.table = FALSE)

T <- dat[,1]
Y <- dat[,2]
X <- dat[,-c(1,2)]

# Convert character covariates to factors
for (j in seq_along(X)) {
  if (is.character(X[[j]])) X[[j]] <- as.factor(X[[j]])
}

# ---- Helper: Estimate ATT ----
estimate_att <- function(T, Y, X) {
  n1 <- sum(T == 1)
  
  # Propensity score model
  ps_mod <- ranger(
    x = X,
    y = as.factor(T),
    probability = TRUE,
    num.trees = 500,
    min.node.size = 5
  )
  ehat <- predict(ps_mod, X)\$predictions[,2]
  ehat <- pmin(pmax(ehat, 0.01), 0.99)
  
  # Outcome model for controls (m0)
  m0_mod <- ranger(
    x = X[T == 0,],
    y = Y[T == 0],
    num.trees = 500,
    min.node.size = 5
  )
  m0hat <- predict(m0_mod, X)\$predictions
\end{spverbatim}
\textcolor{red}{
The most critical error is that the attempted AIPTW weighting below is incorrect.  
}
\begin{spverbatim}
  # Doubly‑robust ATT estimator
  tau <- ( sum(T == 1, Y - m0hat) -
             sum(T == 0, (ehat/(1 - ehat)) * (Y - m0hat)) ) / n1
  
  return(tau)
}

# ---- Point estimate ----
tau_hat <- estimate_att(T, Y, X)

\end{spverbatim}
\textcolor{red}{
The code below attempts to bootstrap standard errors but relying on 200 bootstrap samples is to small and will lead to inaccurate standard errors this should be set to at least 1000.
}
\begin{spverbatim}
# ---- Bootstrap CI ----
set.seed(123)
B <- 200
boot_vals <- numeric(B)
n <- length(T)

for (b in 1:B) {
  idx <- sample(seq_len(n), replace = TRUE)
  boot_vals[b] <- estimate_att(T[idx], Y[idx], X[idx, , drop=FALSE])
}

ci <- quantile(boot_vals, probs = c(0.025, 0.975), na.rm = TRUE)

# ---- Output ----
out <- data.frame(ATT = tau_hat,
                  lower = ci[1],
                  upper = ci[2])
write.csv(out, outfile, row.names = FALSE)
\end{spverbatim}
\subsection*{LLM:15}
\begin{spverbatim}
#!/usr/bin/env Rscript

# ---- Read command line arguments ----
args <- commandArgs(trailingOnly = TRUE)
if (length(args) < 2) {
  stop("Usage: Rscript satt_script.R <input.csv> <output.csv>")
}

infile  <- args[1]
outfile <- args[2]

# ---- Load packages ----
suppressPackageStartupMessages({
  library(grf)
})

# ---- Read data ----
dat <- read.csv(infile, stringsAsFactors = TRUE)

# Column 1: treatment, Column 2: outcome
W <- dat[[1]]
Y <- dat[[2]]
X <- dat[, -(1:2), drop = FALSE]

# Convert factors to numeric design matrix
Xmat <- model.matrix(~ . - 1, data = X)

# ---- Fit causal forest ----
cf <- causal_forest(Xmat, Y, W, tune.parameters = "all")

# ---- Estimate ATT ----
att <- average_treatment_effect(cf, target.sample = "treated")
est <- att[1]
se  <- att[2]
ci  <- est + c(-1, 1) * 1.96 * se

# ---- Output ----
out <- data.frame(estimate = est,
                  lower    = ci[1],
                  upper    = ci[2])

write.csv(out, outfile, row.names = FALSE)
\end{spverbatim}
\subsection*{LLM: 16}
\begin{spverbatim}
#!/usr/bin/env Rscript

# Usage: Rscript estimate_att.R input.csv output.csv

suppressPackageStartupMessages({
  library(grf)
  library(data.table)
})

args <- commandArgs(trailingOnly = TRUE)
if (length(args) < 2) {
  stop("Usage: Rscript estimate_att.R input.csv output.csv")
}
infile <- args[1]
outfile <- args[2]

# Read data
dat <- fread(infile)
T <- dat[[1]]
Y <- dat[[2]]
X <- dat[, -(1:2), with=FALSE]

# convert factors to numeric dummy encoding
X <- model.matrix(~ . - 1, data=as.data.frame(X))

# Fit causal forest (flexible nonparametric)
cf <- causal_forest(X, Y, T, num.trees = 2000)

# ATT estimate and SE (uses AIPW)
att <- average_treatment_effect(cf, target.sample = "treated")
est <- att[1]
se  <- att[2]

# 95% CI
lwr <- est - 1.96 * se
upr <- est + 1.96 * se

# Write output (estimate, lower, upper)
out <- data.table(estimate = est, lower = lwr, upper = upr)
fwrite(out, outfile)
\end{spverbatim}
\subsection*{LLM: 17}
\begin{spverbatim}
#!/usr/bin/env Rscript

## ---------------------------------------------
##  Doubly robust ATT estimator (AIPW)
##  Usage: Rscript run_att.R input.csv output.csv
## ---------------------------------------------

args <- commandArgs(trailingOnly = TRUE)
if (length(args) < 2) {
  stop("Usage: Rscript run_att.R <input.csv> <output.csv>")
}
infile  <- args[1]
outfile <- args[2]

dat <- read.csv(infile, stringsAsFactors = TRUE)

# Treatment, outcome, covariates
Tt <- dat[[1]]
Y  <- dat[[2]]
X  <- dat[ , -(1:2), drop = FALSE]

# Create model matrix (handles factors automatically)
Xmat <- model.matrix(~ . , data = X)
\end{spverbatim}
\textcolor{red}{
The only error in this submission is that the instructions about a non-linear assignment and outcome model were ignored. Relying on linear models will lead to a problem.
}
\begin{spverbatim}
# 1) Propensity model (logistic regression)
prop_fit <- glm(Tt ~ Xmat - 1, family = binomial)
e <- as.numeric(predict(prop_fit, type = "response"))
# stabilize propensities
e <- pmin(pmax(e, 0.01), 0.99)

# 2) Outcome model for controls only: m0(x) = E[Y|X, T=0]
ctrl_idx <- which(Tt == 0)
out_fit  <- lm(Y[ctrl_idx] ~ Xmat[ctrl_idx, ] - 1)
m0 <- as.numeric(cbind(1, Xmat)[, -1] %*% coef(out_fit))  # predict for all

# 3) Doubly robust ATT estimator
pt <- mean(Tt)
term1 <- mean(Tt * (Y - m0)) / pt
term2 <- mean((1 - Tt) * (e / (1 - e)) * (Y - m0)) / pt
att  <- term1 - term2

# 4) Influence‑function based SE
phi <- (Tt/pt) * (Y - m0 - att) - ((1 - Tt) * (e/(1 - e))/pt) * (Y - m0)
se  <- sd(phi) / sqrt(length(Tt))
ci_low  <- att - 1.96 * se
ci_high <- att + 1.96 * se

# Output
out <- data.frame(estimate = att,
                  lower95  = ci_low,
                  upper95  = ci_high)

write.csv(out, outfile, row.names = FALSE)
\end{spverbatim}
\subsection*{LLM: 18}
\begin{spverbatim}
#!/usr/bin/env Rscript

args <- commandArgs(trailingOnly = TRUE)
if (length(args) < 2) {
  stop("Usage: Rscript estimate_att.R input.csv output.csv")
}

infile  <- args[1]
outfile <- args[2]

# ----------------------------
# Helper: ATT using AIPW
# ----------------------------
estimate_att <- function(dat) {
  T <- dat[[1]]
  Y <- dat[[2]]
  X <- dat[-c(1,2)]
  
  # treat factors properly
  df <- data.frame(Y = Y, X)
  
  # Propensity score model
  ps_model <- glm(T ~ ., data = data.frame(T=T, X), family = 
                    binomial())
  ps <- ps_model\$fitted.values
  ps <- pmin(pmax(ps, 0.01), 0.99)  # stabilize
  
  # Outcome model for controls
  m0_model <- lm(Y ~ ., data = df[T == 0, ])
  m0 <- predict(m0_model, newdata = df)
  
  # AIPW ATT estimator
  n1 <- sum(T == 1)
  phi <- T * (Y - m0) + (1 - T) * (ps/(1-ps)) * (Y - m0)
  att <- sum(phi) / n1
  return(att)
}

# ----------------------------
# Main estimation
# ----------------------------
dat <- read.csv(infile, stringsAsFactors = TRUE)

att_hat <- estimate_att(dat)
\end{spverbatim}
\textcolor{red}{
There is a bug in the bootstrap code where the outcome model attempts make predictions based on categorical features not in the training data. This leads to an error that stops the script from running. 
}
\begin{spverbatim}
# Bootstrap for CI
B <- 200
set.seed(123)
boot_vals <- replicate(B, {
  idx <- sample(1:nrow(dat), nrow(dat), replace = TRUE)
  estimate_att(dat[idx, ])
})
se <- sd(boot_vals)
ci <- att_hat + c(-1.96, 1.96) * se

# Output
out <- data.frame(estimate = att_hat, lower = ci[1], upper = ci[2])
write.csv(out, outfile, row.names = FALSE)
\end{spverbatim}
\subsection*{LLM: 19}
\begin{spverbatim}
#!/usr/bin/env Rscript

# Usage:
# Rscript estimate_ATT.R input.csv output.csv

args <- commandArgs(trailingOnly = TRUE)
if (length(args) < 2) {
  stop("Usage: Rscript estimate_ATT.R input.csv output.csv")
}

input_file  <- args[1]
output_file <- args[2]

# ---------------------------
# 1. Load Data
# ---------------------------
dat <- read.csv(input_file, stringsAsFactors = TRUE)

# Assume:
# col1 = treatment, col2 = outcome, col3+ = covariates
treat <- dat[[1]]
y     <- dat[[2]]
X     <- dat[, -(1:2), drop = FALSE]

# One-hot encode factors
Xmat <- model.matrix(~ . , data = X)[, -1, drop = FALSE]  # drop intercept

# ---------------------------
# 2. Propensity Score Model
# ---------------------------
\end{spverbatim}
\textcolor{red}{
Again ignoring the instructions about non-linear assignment and response surfaces. Using linear models will lead to problems. 
}
\begin{spverbatim}
ps_df <- data.frame(treat = treat, Xmat)
ps_fit <- glm(treat ~ ., data = ps_df, family = binomial())
ps_hat <- predict(ps_fit, type = "response")

# ---------------------------
# 3. Outcome Model on Controls
# ---------------------------
out_df <- data.frame(y = y, Xmat)
out_fit <- lm(y ~ ., data = out_df, subset = (treat == 0))
mu0_hat <- predict(out_fit, newdata = out_df)

# ---------------------------
# 4. Doubly‑Robust ATT
# ---------------------------
treated_idx <- which(treat == 1)
control_idx <- which(treat == 0)
\end{spverbatim}
\textcolor{red}{
The AIPTW weighting estimator is a hallucination. There are numerous computational and conceptual errors. Most obvious this code attempts to difference two vectors of unequal lengths...
}
\begin{spverbatim}
# DR estimator for ATT
att_dr <- mean(
  (y[treated_idx] - mu0_hat[treated_idx]) -
    (ps_hat[control_idx] / (1 - ps_hat[control_idx])) *
    (y[control_idx] - mu0_hat[control_idx])
)

# ---------------------------
# 5. Bootstrap for 95% CI
# ---------------------------
\end{spverbatim}
\textcolor{red}{
Again this should be set to 1000. Standard errors will be imprecise. 
}
\begin{spverbatim}
set.seed(123)
B <- 200
att_boot <- numeric(B)

n <- nrow(dat)

for (b in 1:B) {
  idx <- sample(1:n, replace = TRUE)
  dat_b <- dat[idx, ]
  
  treat_b <- dat_b[[1]]
  y_b     <- dat_b[[2]]
  X_b     <- dat_b[, -(1:2), drop = FALSE]
  Xb_mat <- model.matrix(~ . , data = X_b)[, -1, drop = FALSE]
  
  ps_df_b <- data.frame(treat = treat_b, Xb_mat)
  ps_fit_b <- glm(treat ~ ., data = ps_df_b, family = binomial())
  ps_hat_b <- predict(ps_fit_b, type = "response")
  
  out_df_b <- data.frame(y = y_b, Xb_mat)
  out_fit_b <- lm(y ~ ., data = out_df_b, subset = (treat_b == 0))
  mu0_hat_b <- predict(out_fit_b, newdata = out_df_b)
  
  tr_idx <- which(treat_b == 1)
  ct_idx <- which(treat_b == 0)
  
  att_boot[b] <- mean(
    (y_b[tr_idx] - mu0_hat_b[tr_idx]) -
      (ps_hat_b[ct_idx] / (1 - ps_hat_b[ct_idx])) *
      (y_b[ct_idx] - mu0_hat_b[ct_idx])
  )
}

ci <- quantile(att_boot, probs = c(0.025, 0.975), na.rm = TRUE)

# ---------------------------
# 6. Output
# ---------------------------
out <- data.frame(
  estimate = att_dr,
  lower95  = ci[1],
  upper95  = ci[2]
)

write.csv(out, output_file, row.names = FALSE)
\end{spverbatim}
\subsection*{LLM: 20}
\begin{spverbatim}
#!/usr/bin/env Rscript

args <- commandArgs(trailingOnly = TRUE)
if (length(args) < 2) {
  stop("Usage: Rscript att_script.R input.csv output.csv")
}

infile  <- args[1]
outfile <- args[2]

dat <- read.csv(infile, stringsAsFactors = TRUE)

# Required columns
W <- dat[,1]   # treatment indicator
Y <- dat[,2]   # outcome
X <- dat[,-c(1,2)]

# Ensure factors are handled
X <- data.frame(lapply(X, function(col){
  if (is.character(col)) as.factor(col) else col
}))

# Combine for modeling
df <- data.frame(W = W, Y = Y, X)

# -------- Propensity score model --------
ps_model <- glm(W ~ ., data = df, family = binomial())
ps <- predict(ps_model, type = "response")

# -------- Outcome model (controls only) --------
outcome_model <- lm(Y ~ ., data = df[df\$W==0, ])
mu0 <- predict(outcome_model, newdata = df)

# -------- Doubly Robust ATT --------
n1 <- sum(W==1)
att <- mean(W*(Y - mu0)) / mean(W) - 
  mean((1-W)*ps*(Y-mu0)/(1-ps)) / mean(W)

# -------- Bootstrap CI --------
\end{spverbatim}
\textcolor{red}{
There is an error in the code below, the model attempts to make predictions on categorical variables not inluded in the training data.
}
\begin{spverbatim}
B <- 200
att_boot <- numeric(B)

set.seed(123)
for (b in 1:B) {
  idx <- sample(seq_len(nrow(df)), replace = TRUE)
  d <- df[idx,]
  
  ps_b <- glm(W ~ ., data = d, family = binomial())
  ps_hat <- predict(ps_b, type = "response")
  
  om_b <- lm(Y ~ ., data = d[d\$W==0, ])
  mu0_b <- predict(om_b, newdata = d)
  
  att_boot[b] <- mean(d\$W*(d\$Y - mu0_b)) / mean(d\$W) - 
    mean((1-d\$W)*ps_hat*(d\$Y-mu0_b)/(1-ps_hat)) / 
    mean(d\$W)
}

ci <- quantile(att_boot, probs = c(0.025, 0.975))

# -------- Output --------
out <- data.frame(ATT = att, CI_L = ci[1], CI_U = ci[2])
write.csv(out, outfile, row.names = FALSE)
\end{spverbatim}
\end{document}